\documentclass[twocolumn,usenatbib,useAMS,usegraphicx]{mn2e}

\usepackage{amsmath,amssymb}
\usepackage{aas_macros}
\usepackage{multirow,multicol}
\usepackage{xspace}
\usepackage{url}

\newcommand{\Msun}{{\rm M_\odot}}
\newcommand{\dd}{{\rm {d}}}
\newcommand{\EPS}{{\small EPS}\xspace}
\newcommand{\WHIM}{{\small WHIM}\xspace}
\newcommand{\AGN}{{\small AGN}\xspace}
\newcommand{\IGM}{{\small IGM}\xspace}
\newcommand{\phn}{\phantom{0}}%
\newcommand{\etam}{\eta_{\rm min}}
\newcommand{\rhob}{\rho_{\rm b}}

\newcommand{\omb}{\Omega_{\rm b}}
\newcommand{\omdm}{\Omega_{\rm dm}}
\newcommand{\mres}{M_{\rm res}}
\newcommand{\Nsims}{{100}\xspace}

\title{Balancing the Baryon Budget: The fraction of the IGM due to Galaxy Mergers}

\author[Manodeep Sinha, Kelly Holley-Bockelmann]
{Manodeep Sinha$^{1}$\thanks{E-mail:~manodeep.sinha@vanderbilt.edu, k.holley@vanderbilt.edu}
, Kelly Holley-Bockelmann$^1$ \\
$^1$Department of Physics \& Astronomy, Vanderbilt University  \\
}

\begin{document}

\maketitle

\begin{abstract}
Observations indicate that roughly 60\% of the baryons may exist in a Warm-Hot Intergalactic 
Medium (\WHIM) at low redshifts. Following up on previous results showing that gas is released 
through galaxy mergers, we use a semi-analytic technique to estimate the fraction of gas mass
lost from haloes solely due to mergers. We find that up to $\sim$ 25\% of the
gas in a halo can unbind over the course of galaxy assembly. This process does 
not act preferentially on smaller mass haloes; bigger haloes {\emph {always}} release larger 
amounts of gas in a given volume of the Universe. However, if we include multi-phase
gas accretion onto haloes, we find that only a few percent is unbound.
We conclude that either non-gravitational processes may be in play to heat up the gas in 
the galaxies prior to unbinding by mergers or most of the baryons in the \WHIM have never fallen into 
virialised dark matter haloes. We present a budget for stocking the \WHIM compiled from recent
work.
\end{abstract}

\begin{keywords}

galaxies: evolution - galaxies: haloes - galaxies: interactions - intergalactic medium - 
methods: numerical - cosmology: large-scale structure

\end{keywords}

\section{Introduction}
The baryon budget shows significant evolution from $z \sim 3$, and results in
an apparent baryon deficit today \citep{FHP98,FP04}. At
high redshift, most of the baryonic mass is in the Ly-$\alpha$ forest \citep{FHP98,FP04},
while at low redshifts over half of the baryons are as yet undetected.
The consensus is that the majority of the `missing'
baryons are actually in regions of low overdensity, $\delta\rho/\rho \sim 10-100 $ 
\citep[e.g][]{CO99,DHKW99,DC01,KRCS05,CO06,DM06,DO07} with
temperatures in the range $10^5-10^7$ K -- commonly referred to as the Warm-Hot Intergalactic Medium (\WHIM).

The immediate question is: how is this \WHIM produced~?
Some form of mass and  energy injection is essential to create this hot reservoir of gas;
this form of feedback must both regulate the gas in galaxies and 
the metal content of the Intergalactic Medium (\IGM). There has been much 
numerical work to incorporate various feedback mechanisms in
an attempt to solve this puzzle \citep[e.g.,][and references therein]{CO99,NS01,DC01,KRCS05,CO06,DO07}.
Cosmological simulations seem to suggest that gravitational collapse during galaxy formation can produce
and maintain the majority of the \WHIM at $10^5 - 10^7$ K \citep{CO99,DHKW99,CO06,CDM01,DO07}. 
Supernova feedback provides another avenue to generate the \WHIM; 
for star bursts of 100 $M_\odot$ per year, as much as $20\%$ of the hot gas in a Milky Way mass galaxy 
can be unbound \citep{STWS06,KSW07}. However,  SNe feedback may be a self regulating process, in that a starburst 
also heats the remaining gas and may damp the star formation rate, which in turn 
would reduce the fraction of unbound gas \citep[e.g.][]{STWS08}. Combining these effects, it is commonly thought 
that galaxies with host halo mass $\gtrsim 10^{11} $ M$_{\odot}$ lose $\lesssim 10\%$ of their 
gas through SN feedback, while low mass haloes may be entirely depleted of gas by this mechanism\citep{YK97,MF99,E00,STWS06}. 
A third possibility is that the radiation from an accreting supermassive black hole 
could power large-scale winds to blow mass out of the galaxy \citep[see][]{SO04,MQT05,HH05,HHM05,HH06,CS06,SSMH07}. 
For a fixed amount of energy, all the non-gravitational feedback mechanisms are more effective
in low mass galaxies due to their shallower potential. However, observations suggest that low-mass galaxies are in general 
more gas-rich and are less likely to have suffered a gas blow-out \citep{K04,GBMW06}.  

In \citet[][hereafter, SH09]{SH09}, we show that hot gas is driven  into the \IGM by galaxy mergers.
The amount of hot halo gas lost depends strongly on the energy of the merger; 
it is possible for low mass galaxies to retain their gas in this scenario during low-speed
or distant encounters. However, SH09 only estimated the mass lost during a single
merger. When all the mergers in the Universe are considered, this could heat and drive a significant
portion of the total baryon budget into the \WHIM. In principle, this process could join \AGN and 
star formation feedback as a way to populate the \WHIM, and we find that this method operates
preferentially in {\em more} massive haloes. To estimate the total fraction of gas released by mergers, we construct a series of 
analytic halo merger trees using a publicly available\footnotemark semi-analytic Extended Press-Schechter (\EPS) code \citep{PCH08}.
\footnotetext{\url{http://star-www.dur.ac.uk/~cole/merger_trees/}}

In Section~\ref{section:eps} we describe the theory of halo merger trees, in Section~\ref{section:method}
we outline the experiments designed to track the gas ejected via galaxy mergers, in Section~\ref{section:results}
we present the results for the halo gas ejected by this process and Section~\ref{section:discussion} contains
the discussion. 

\section{Constructing merger-trees}\label{section:eps}
Observations reveal that we live in low-density, $\Lambda$-dominated flat Universe
\citep{RF98,PG97,SB07,K08}. In such a Universe,
haloes form hierarchically, with smaller haloes  forming early on and merging into
larger structures at later times. This process of halo formation is dictated 
by gravitational processes, and 
an analytic formalism yields the number density of haloes as a function of mass and 
redshift \citep{PS74}. 
However, this does not constrain the merger rates for any given halo as a function
of redshift. To this end, the Press-Schechter formalism has been extended to 
calculate a merger history of a halo in the form of a binary merger tree \citep{LC93,B91,BCEK91}.
These merger trees are computationally much less expensive than an N-body simulation, and are widely used to explore
and constrain theories of galaxy evolution, black hole growth, etc. We use the technique here to estimate
the gas lost to the \WHIM via galaxy mergers. 

In the \EPS model of \citet{PCH08}, the conditional mass function $f(M_1|M_2)$
gives the fraction of mass from a halo with mass $M_2$ at a redshift $z_2$
that was contained in a progenitor halo of mass $M_1$ at a previous redshift $z_1$:
\begin{eqnarray}
\lefteqn{\displaystyle{
f(M_1 \vert M_2)\, d\ln M_1 =
\sqrt{\frac{2}{\pi}} \,
\frac{\sigma_1^2
  (\delta_1-\delta_2)}{[\sigma_1^2-\sigma_2^2]^{3/2}} } \, \times} &&
\nonumber \\
&&\displaystyle{ \exp\left[ - \frac{1}{2} \frac{(\delta_1-\delta_2)^2}{(\sigma_1^2-\sigma_2^2)}\right]
\left\vert \frac{d\ln\sigma}{d\ln M_1} \right\vert \,
d\ln M_1}\, ,
\label{eqn:eps}
\end{eqnarray}
where $\delta_1$ and $\delta_2$ represent linear overdensities for collapse at  redshifts $z_1$
and $z_2$ and $\sigma \equiv \sigma (M)$. The derivative of this equation under the limit $z_1 \rightarrow z_2$
yields the number $N$ of progenitors of mass $M_1$  that make up a halo of mass $M_2$ for a small step in
redshift space of $\dd z_1$. This is written as:
\begin{equation}
\frac{dN}{d M_1} =  \frac{1}{M_1} \ \frac{df(M_1 \vert M_2)}{dz_1} \frac{M_2}{M_1} dz_1
\qquad (M_1 < M_2) .
\label{eqn:dnp}
\end{equation}
Specifying a minimum mass resolution $\mres$ allows us to compute the mean number of progenitors $N_P$
with mass $M_1$ in a mass range $\mres < M_1 < M_2/2$ via the following equation:
\begin{equation}
N_P=\int_{\mres}^{M_2/2} \frac{dN}{d M_1} \ dM_1\, .
\label{eqn:progenitors}
\end{equation}
The mass fraction $F$ of the final halo $M_2$ that is accreted below $\mres$ can be
estimated from:
\begin{equation}
F= \int_{0}^{\mres} \frac{dN}{d M_1}\ \frac{M_1}{M_2} \
dM_1\,.
\label{eqn:massfrac}
\end{equation}
A binary merger tree can then be constructed given $M_2$ and $z_2$. We used this technique
to construct a set of twelve merger-trees, which we outline in Section~\ref{section:method}.
We have assumed a flat $\Lambda$CDM cosmology with $\omb = 0.044,\, \omdm = 0.214,\,\Omega_\Lambda = 0.742, \,
\sigma_8 = 0.796 \, {\rm and} \, h = 0.719$, consistent with the {\small WMAP} 5-year cosmology parameters \citep{K08}.

\section{Method}\label{section:method}
As shown in SH09, the amount of gas\footnotemark released by a galaxy merger depends on the mass ratio and the original 
gas content of the haloes. 
\footnotetext{Since we do not model star formation in our semi-analytic approach, we will use the terms gas and baryons interchangeably.}
To incorporate this effect within a merger tree we take the following approach:
we seed each halo with a gas fraction ($f_{\rm seed}$) and assume a galaxy merger with a mass ratio greater than $\etam$ 
unbinds a fraction of this gas ($f_{\rm unb}$). We also assume that as the halo grows by diffuse accretion from the
\IGM, it also accretes gas at the universal gas fraction, increasing the halo gas content.
Recent simulations \citep[see][]{KKWD05,K09a,DB09} show that gas does not necessarily heat up to the halo virial temperatures; the 
majority of the haloes at low-$z$ are only accreting cold gas. To estimate the effect of this multiphase
accretion models, we divided the halo gas mass into hot and cold components in accordance with Figure 3 of \citet{K09a}.
After this partitioning, we follow the same procedure, except now we only unbind
gas from the hot gas component. Table~\ref{table:models} outlines the parameters for the twelve experiments.

We designed these twelve experiments to bracket the likely effect that galaxy mergers have on populating
the \WHIM. A reasonable upper limit is set by allowing even minor mergers ($\etam = 0.1$) to unbind a fixed
fraction ($f_{\rm unb} = 0.1$) of the progenitor gas mass (run Minor1). Our lower limit is set by seeding only the massive
haloes ($M_{\rm halo} \geq M_{\rm min} = 10^{10}\,\Msun$) with gas at the universal gas fraction and allowing only 
major mergers ($\eta = 0.3$) to release gas (run Major5). SH09 found that roughly
equal-mass mergers can release up to 20\% of their initial gas mass, and since the merger rate (per halo per redshift)
is relatively flat from $0.3<\eta<1.0$ \citep[see Fig.~8 in][]{FM08}, we argue that $\etam = 0.3$, $f_{\rm unb} = 0.1$ 
is a good average scenario. Haloes more massive than $10^{13}\,\Msun$, representing groups or clusters of galaxies, can not be 
faithfully modelled using this binary galaxy merger mechanism and have been left out. 

The input parameters are the final halo mass, $M_2$, the initial redshift, $z_1=10$ and the mass resolution, $\mres$. We
explore a range of final halo masses from $M_2 = 10^8 - 10^{13}\,\Msun$.
We use 100 logarithmically spaced mass bins to create a merger tree for a specific $M_2$ at the present epoch. 
To account for cosmic variance, we run \Nsims realisations of a 
fixed halo mass. Thus, overall we create 100 present day halo samples 
with \Nsims realisations for a fair sample of possible hierarchical 
merger histories of structure in the Universe. For each merger tree, we set $\mres \,=\, 
M_2\times10^{-5}$. For  $10^{11}\,\Msun$ haloes, this value of $\mres$ is comparable to the mass of an individual 
dark matter particle in our numerical simulations (SH09). We tested the effect of changing 
$\mres$, $z_1$ and the number of redshift levels and found that our choices
produce convergent results for the estimation of the unbound gas. 

With these merger histories, we follow all mergers from $z=10$ to $z=0$
that lead to a halo of mass $M_2$, and eject a fraction of gas from the mergers 
with mass ratios greater than $\etam$.
The cumulative sum of the unbound gas produces the total gas released in assembling a particular halo. 
This yields the fractional gas lost by $z=0$ on a {\emph {per halo}} basis.
We repeat this process for \Nsims realisations, which provides
the variance in the gas lost. We can find the total gas released in generating all haloes 
in the Universe by convolving with the co-moving number density of those haloes at 
$z=0$\citep{WAHT06}. Summing over the final halo masses yields the effect of halo assembly 
on populating the \WHIM.

\begin{table}
\caption{The initial parameters for the twelve merger trees. 
Column 1 is the minimum merger ratio for gas to get ejected from haloes, column 3 is
the fraction of the halo mass used to seed newly-appeared haloes (equal
to $\omb/\omdm$ for all experiments other than Major3 and Minor3), column 4 is
the fraction of the halo gas that unbinds during a merger, column 5 
is the minimum halo mass that can retain gas and column 6 shows the entire mass range of haloes
for which merger trees were made.}
\centering
\begin{tabular}{cccccr}
\hline \hline \\[-2ex]
   \multicolumn{1}{c}{$\mathbf{\etam}$} &
   \multicolumn{1}{c}{\textbf{Run}} &
   \multicolumn{1}{c}{$\mathbf{f_{\rm seed}}$} &
   \multicolumn{1}{c}{$\mathbf{f_{\rm unb}}$} &
   \multicolumn{1}{c}{$\mathbf{M_{\rm min}}$} &
   \multicolumn{1}{c}{\textbf{Mass range}} \\\hline \\[-2ex]
   \multicolumn{1}{c}{\textbf{[-]}} &
   \multicolumn{1}{c}{\textbf{[-]}} &
   \multicolumn{1}{c}{\textbf{[-]}} &
   \multicolumn{1}{c}{$\mathbf{[\%]}$} &
   \multicolumn{1}{c}{$\mathbf{[\log \Msun]}$} &
   \multicolumn{1}{c}{$\mathbf{[\log \Msun]}$} \\\hline \hline \\[-2ex]

\multirow{6}{*}{0.33} 
& Major1      & 0.21  & 10.0   & -     & 8.0 - 13.0  \\
& Major2      & 0.21  & random & -     & 8.0 - 13.0  \\
& Major3      & random& 10.0   & -     & 8.0 - 13.0  \\
& Major4      & 0.21  & 10.0   & -     & 10.0 - 13.0  \\
& Major5      & 0.21  & 10.0   & 10.0  & 10.0 - 13.0  \\
& Major-Keres & 0.21  & 10.0   & -     & 8.0 - 13.0  \\[0.7ex] \hline
\multirow{6}{*}{0.10} 
& Minor1      & 0.21   & 10.0   & -    & 8.0 - 13.0  \\
& Minor2      & 0.21   & random & -    & 8.0 - 13.0  \\
& Minor3      & random & 10.0   & -    & 8.0 - 13.0  \\
& Minor4      & 0.21   & 10.0   & -    & 10.0 - 13.0  \\
& Minor5      & 0.21   & 10.0   & 10.0 & 10.0 - 13.0\\
& Minor-Keres & 0.21   & 10.0   & -    & 8.0 - 13.0\\[0.7ex] 
\hline\hline 
\end{tabular}
\label{table:models}
\end{table}

\section{Results}\label{section:results}
\begin{figure}
\includegraphics[scale=0.47]{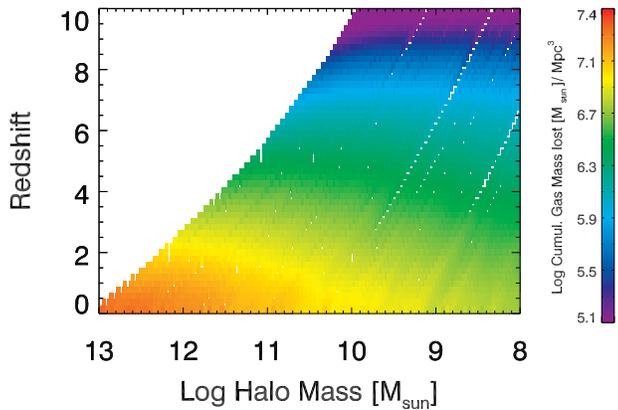}%
\caption{\small The cumulative gas mass lost from haloes by mergers in a co-moving Mpc$^3$ volume
as a function of halo mass and redshift for Major1. Despite their larger number density, 
smaller haloes systematically lose less mass than the bigger haloes. The gas mass lost is 
obtained by taking a mean of \Nsims realisations. The pixels reflect the bin size in 
mass and redshift. This figure is summing the unbound gas mass along redshift and multiplying
by the co-moving number density of the corresponding haloes at $z=0$.}
\label{figure:gasmasslostwhalo}
\end{figure}

Figure~\ref{figure:gasmasslostwhalo} shows the redshift evolution of the cumulative gas mass lost from all 
haloes in a co-moving Mpc$^3$ volume for the run Major1. To generate Figure~\ref{figure:gasmasslostwhalo}, 
we first take the mean of \Nsims realisations for the unbound gas mass in each redshift step for each halo. 
This unbound gas mass is added up along the redshift track to yield the cumulative mass at each redshift step
and then multiplied by the co-moving number density of that particular halo at $z=0$. This is the
cumulative co-moving density of the unbound gas for one halo mass. Repeating this process for the 100 final 
halo masses yields the individual tracks spanning the x-axis.
Figure~\ref{figure:gasmasslostwhalo} shows that the most massive haloes unbind the most gas at all redshifts, in spite of their 
lower number densities. For example, the current number density in a co-moving 
Mpc$^3$ of a $10^{13}\, \Msun$ halo  is $\sim 10^6$ times smaller than for a $10^8\,\Msun$ halo;
so the mass-density of the $10^8 \,\Msun$ halo is an order of magnitude larger than the $10^{13}\,\Msun$ halo.
This biasing towards higher mass is explained by the hierarchical assembly of haloes -- 
more massive objects today undergo many more mergers in the past\footnotemark. 

\footnotetext{There may be an additional effect from the higher gravitational potential energy 
($\propto v_{\rm circ}^2 $) involved during mergers of massive galaxies \citep[see][]{JNO09}.}

\begin{figure}
\centering
\includegraphics[scale=0.55]{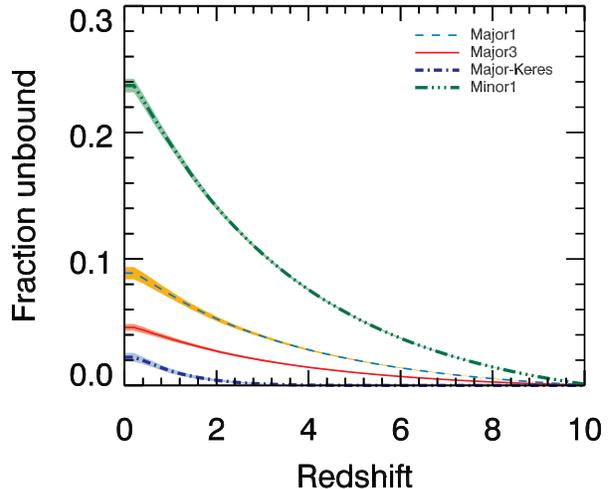}%
\caption{\small The evolution with redshift of the gas mass lost as a fraction of baryon density in the Universe 
baryon content in the merger trees assuming universal gas fraction. If we assume that the trend holds globally, then
the Y-axis can be thought to be normalised by $\rhob$. The shaded region shows the 1-$\sigma$ deviation at each \
redshift from \Nsims realisations. }
\label{figure:gasmasslostwz}
\end{figure}

Figure~\ref{figure:gasmasslostwz} shows the redshift evolution of the unbound gas mass over the total baryon mass 
in all the haloes considered in the merger tree. We find that 9\% and 24\% of the baryons can be ejected
by mergers for the Major1 and Minor1 runs respectively. 
The mass range of $10^{10}-10^{13}\,\Msun$ and $10^{8}-10^{13}\,\Msun$ contain 39\% and 52\% of the
total collapsed mass in the Universe respectively. Thus, the IGM pollution caused by the mergers presented in this 
paper can only reflect the history of at most half the total matter. If we assume that the same pattern holds 
true globally, then the fractions presented here (Figure~\ref{figure:gasmasslostwz}) can be interpreted as normalised by
the total baryonic matter density of the Universe. Notice that the fraction of gas lost increases more rapidly with redshift 
for $\etam=0.1$ -- this is because 10:1 mergers occur more frequently than 3:1 \citep[e.g.,][]{FM08, GGBN09}. 

We can interpret Figure~\ref{figure:gasmasslostwz} in the following way: in the Major1 run, the convergence to 10\%
of the universal gas mass is tantamount to saying that the average halo undergoes one major
merger in a Hubble time, since we set major mergers to release 10\% of the gas mass. Likewise,
the convergence of the Minor1 run can be understood by noting that minor mergers ($\eta > 0.1$)
are  $2-3$ times more frequent than major mergers ($\eta > 0.3$, \citet[see bottom panel of Fig. 8 in][]{FM08}). 
Thus, the overall unbound density converges to $\sim 20-30\%$ for the Minor1 run. 

In run Major-Keres with multiphase accretion, we find that only $\sim$ 2\% of the
gas can be released due to mergers. Since the simulations of SH09 only included hot gas, we
chose to unbind only from that phase. In the multiphase scenario, too much gas
is in the cold phase and hence, can not be released during mergers. Even adding a mechanism to
heat cold gas by major mergers \citep[Eqn. 4][]{CPJS04} does not convert enough cold gas into a hot phase
to be unbound later. If the haloes are only accreting cold gas and this gas can not be 
unbound from the haloes before heating it first, then the gas currently populating the \WHIM may not have 
ever fallen into virialised haloes.  

In a given merger tree, a fraction of unbound gas is released by mergers between small haloes. To
isolate the \WHIM fraction (Table~\ref{table:results}, Column 5)created during the assembly
of {\emph {only}} the massive galaxies, we run two sets of merger trees with a lower mass limit of $10^{10}\,\Msun$. 
Table~\ref{table:results} shows that most of the unbound gas that is released comes during the formation
of the massive galaxies. In particular, Major4, with only the massive haloes, produces nearly all of the 
unbound gas  produced in the Major1 run. 

Although small haloes merging with massive haloes do not eject any gas, these minor accretion
events increase the gas content of the remnant. This could potentially increase the amount 
of gas released by massive haloes in future mergers. 
However, if processes like SN feedback evacuate the gas  from low mass haloes, these low mass haloes  
can not increase the gas content of the massive haloes.
To mimic this effect, we run two sets of merger trees with a lower mass limit of $10^{10}\,\Msun$ and 
only allow haloes larger than $M_{\rm min}$ to contain gas. 
In this scenario (Major5 and Minor5), all small haloes are completely devoid of gas and therefore
do not contribute to the gas mass of the big haloes. With this constraint, we find a \WHIM fraction
of $\sim$ 3\% and 8\% for $\etam = 0.3$ and $0.1$ respectively. 

\begin{table}
\caption{A census of the unbound gas at $z=0$ produced by galaxy mergers. 
Column 2 is the mass of unbound gas in a co-moving Mpc$^3$, 
column 3 is the ratio of unbound gas mass and the expected
universal gas content ($M_{\rm halo} \times \omb/\omdm$).
Column 4 is the ratio of the mean unbound gas in a co-moving Mpc$^3$ to the mean gas 
mass left the haloes. Column 5 shows the fraction of the \WHIM generated; 
we assume that the \WHIM contains 60\% of all the baryons. The gas mass in a co-moving 
Mpc$^3$ volume for the haloes considered here is $7.6\times10^9\,\Msun$.}
\centering
\scalebox{0.85}{
\begin{tabular}{ccccc}
\hline \hline \\[-2ex]
   \multicolumn{1}{c}{\textbf{Run}} &
   \multicolumn{1}{c}{$\mathbf{M_{\rm gas,unb}}$} &
   \multicolumn{1}{c}{$\mathbf{\dfrac{<M_{\rm unb,gas}>}{<M_{\rm gas, univ}>}}$} &
   \multicolumn{1}{c}{$\mathbf{\dfrac{<M_{\rm unb,gas}>}{<M_{\rm gas,gal}>}}$} &
   \multicolumn{1}{c}{$\mathbf{f_{\rm WHIM}}$} \\\hline \\[-2ex]
   \multicolumn{1}{c}{\textbf{[-]}} &
   \multicolumn{1}{c}{$[\mathbf{10^8 \Msun/Mpc^3}]$} &
   \multicolumn{1}{c}{\textbf{[-]}} &
   \multicolumn{1}{c}{\textbf{[-]}} &
   \multicolumn{1}{c}{\textbf{[-]}} \\\hline \hline \\[-2ex]
Major1            & \phn 6.7 $\pm$ 0.4 & 0.09 & 0.10 & 0.15 \\
Major2            & \phn 3.5 $\pm$ 0.2 & 0.05 & 0.05 & 0.08 \\
Major3            & \phn 5.7 $\pm$ 0.3 & 0.07 & 0.09 & 0.12 \\
Major4            & \phn 5.0 $\pm$ 0.3 & 0.09 & 0.10 & 0.15 \\
Major5            & \phn 1.0 $\pm$ 0.1 & 0.02 & 0.04 & 0.03 \\
Major-Keres       & \phn 1.7 $\pm$ 0.3 & 0.02 & 0.03 & 0.03 \\[0.7ex] \hline

Minor1            &     17.9 $\pm$ 0.4 & 0.24 & 0.32 & 0.39 \\
Minor2  	 	  & \phn 9.7 $\pm$ 0.3 & 0.13 & 0.15 & 0.21 \\
Minor3  	 	  &     15.2 $\pm$ 0.4 & 0.20 & 0.29 & 0.33 \\
Minor4  	 	  &     13.4 $\pm$ 0.4 & 0.24 & 0.32 & 0.40 \\
Minor5  	 	  & \phn 2.9 $\pm$ 0.1 & 0.05 & 0.13 & 0.08 \\
Minor-Keres	 	  & \phn 3.8 $\pm$ 0.3 & 0.03 & 0.06 & 0.06 \\[0.7ex] \hline
\hline \\
\end{tabular}
}
\label{table:results}
\end{table}

\section{Discussion}\label{section:discussion}
In this paper we show that a significant portion of the \WHIM can be generated
by gas ejected from galaxies during mergers. Our semi-analytic prescription shows that 
up to $\sim$ 25\% of the gas (assuming universal gas fraction) in 
haloes of mass $10^8 - 10^{13}\, \Msun$ can be ejected by mergers. 
Given an observed gas mass at $z=0$, it is possible to infer the typical gas mass
that was unbound from assembling that halo (column 4, Table~\ref{table:results})
For comparison with SN feedback, a quiescent Milky-Way type halo with star formation rate of 1-10 $\Msun{\rm yr^{-1}}$
would unbind $\leq 2\%$ of the gas content \citep{STWS08}. We also find that multiphase gas accretion
drastically reduces the amount of unbound gas from mergers, down to a few percent of the gas mass.
In contrast with previous numerical work involving non-gravitational feedback, where the effects of 
mass loss are severe in smaller haloes, this merger mechanism unbinds gas preferentially from massive haloes.
There is no selective unbinding of gas from dwarf galaxies, in line with observational evidence 
suggesting that dwarf galaxies are more gas-rich and therefore may not have suffered a gas blow-out \citep{K04,GBMW06}. 

This form of {\it gravitational feedback} may even play a larger role in regulating the stellar mass function: \citet{K09a} 
show that simulated  galaxies exhibit a discrepancy with the observed stellar mass function \citep{B03} for both  
high and the low mass galaxies. They conjecture that the key to solving this discrepancy is through a 
feedback mechanism that works for halos $\gtrsim 10^{12}\, \Msun$ -- akin to our scenario. 
If the merger-ejection process is very efficient, then the current day haloes may be very gas-poor. It is conceivable that 
the current stellar mass {\em and} the gas fraction of a galaxy constrains the mean stellar mass and gas content of the 
galaxies of the past.

Overall, we find that for our most reasonable scenario, $\sim 15\%$ of the \WHIM can be generated
through galaxy mergers. If previous work on large-scale gravitational shocks proves correct \citep{CO99, DC01}, $\sim 66\%$ of the
\WHIM can be attributed to gas that may have never fallen into a halo. In addition, recent studies have shown
that roughly 20\% can be produced via non-gravitational feedback, such as SNe and AGN \citep[e.g.,][]{CO06}. Therefore, 
with these three mechanisms to populate the \WHIM, it may well be true that the baryon budget is balanced.

\section{Acknowledgements}
We thank the {\small GALFORM} team for making the {\small EPS} code publicly available. We thank Andreas Berlind
for sharing his halo mass function code. This work was conducted in part using 
the resources of the Advanced Computing Center for Research and Education at Vanderbilt University, Nashville, TN. 
We also acknowledge support from the NASA grant NNX08AG74G and from the NSF Career award AST-0847696. 

\bibliographystyle{apj}
\bibliography{Biblio-Database}

\end{document}